\documentclass[twocolumn,showpacs,aps,prl]{revtex4}
\usepackage{graphicx}
[1999/03/29 PSNFSS v.7.2
Times font as default roman
: S Rahtz]

\begin{document}
\renewcommand{\thefootnote}{\fnsymbol{footnote}}
\sloppy
\newcommand{\rp}{\right)}
\newcommand{\lp}{\left(}
\newcommand \be  {\begin{equation}}
\newcommand \ba {\begin{eqnarray}}
\newcommand \ee  {\end{equation}}
\newcommand \ea {\end{eqnarray}}

\title{A Mechanism for Pockets of Predictability in Complex Adaptive Systems} 
\thispagestyle{empty}

\author{J\o rgen Vitting Andersen$^{1,2}$ and Didier Sornette$^{2,3}$}
\affiliation{$^1$ U.F.R. de Sciences Economiques, Gestion, Math\'ematiques 
et Informatique, CNRS UMR 7536 and Universit\'e Paris X-Nanterre, 92001
Nanterre Cedex}
\affiliation{$^2$ Laboratoire de Physique de la Mati\`ere Condens\'ee,
CNRS UMR 6622 and Universit\'e de Nice-Sophia Antipolis, 06108
Nice Cedex 2, France}
\affiliation{$^3$ Institute of Geophysics and Planetary Physics
and Department of Earth and Space Sciences,
University of California, Los Angeles, CA 90095}

\email{vitting@unice.fr, sornette@moho.ess.ucla.edu}

\date{\today}

\begin{abstract}

We document a mechanism operating in complex adaptive systems
leading to dynamical pockets of predictability (``prediction days''),
in which agents collectively take
predetermined courses of action, transiently decoupled from past history.
We demonstrate and test it out-of-sample on synthetic minority and
majority games as well as on real financial time series. The
surprising large frequency of these prediction days implies a collective
organization of agents and of their strategies which condense into
transitional herding regimes.

\end{abstract}


\maketitle

The concept of complex adaptive feedbacks is increasingly used
to understand the earth climate, immune systems,
nervous systems, multicellular organisms, insect
societies, ecologies, economies, human societies, stock markets,
distributed computing systems, large-scale communication networks.
A trademark of such systems is the occurrence of extreme events, which
are in general believed to be unpredictable \cite{Karplus}. 
However, a few recent works suggest a degree of predictability
in some cases, for instance associated with transient
herding phases in the population of agents \cite{JSL}, or signaled
by empirical patterns \cite{Lo}. Previous attempts to 
examine the predictability of large future changes have used models of
evolving agents competing for a limited resource \cite{Farmer,Lamper}. These
agents have memory and use complex (possibly learning) algorithms.
Although they can show predictive power, the
complexity of agent-based models has not provided an understanding
of the factors that lead to the predictions except to say that they are a
consequence of the information incoded in the 
system's global state \cite{Lamper}. This is particularly 
problematic for concrete applications for which predictions become
credible and usable only when based upon a sound physical understanding of their
limitations and of the range and likelihood of competing scenarios.

Let us consider a generic
multi-agent system comprising a population of $N$ agents of which less
than half of the agents can win at each time step. Each agent will
thus seek to be in the minority group. In \cite{Cavagna}, it is was
shown that, for such Minority Games (MG), it is the information common
to all the agents, rather than the feedback of their actions onto the
price, which determines the dynamics of the game. Here, we extend this
observation and show the surprising fact that, for minority as well as
for majority games \cite{$-game}, at certain times, the information
contained in a few last moves becomes irrelevant in the agents' decision
making, a situation that we refer to as ``decoupling.'' This leads to
dynamical pockets of predictability, with agents taking a predetermined
course of action which is decoupled from the price history of the system
over a finite number of time steps. We are able to unveil the origin of
these pockets of predictability that we test out-of-sample on synthetic
as well as on real financial time series using a suitably calibrated
agent-based model.

Our results apply to systems which can be represented
by agent-based models, in which each agent $i$ has a finite memory over
$m_i$ time steps fed to his $s_i$ strategies. Without
significant loss of generality, we follow most previous models \cite{BookMG}
and assume that $m_i=m$ and $s_i=s$ for all agents.
At each time step $t$, each of the $N$ agents makes either a decision $a^{\mu_m}_i (t) = \pm 1$
(yes or no, itinary A or B, buy or sell, etc.) 
using  his best (among the $s$) strategy or does nothing, based on the 
available information of the last $m$ time steps. The available
common information $\mu_m(t)$ at time $t$
is the series of past total actions of all the agents over the last $m$ time steps
\be
\mu_m(t) = \{{1+{\rm sign}[A^{\mu} (k)] \over 2}; k = t-m+1, ..., t\}~,
\ee
where $A^{\mu} (k) = \sum_i a^{\mu}_i (k)$. Thus, $\mu_m(t)$ is a string 
of $m$ binary digits $0$ (a majority of agents played $-1$) or $1$ 
(a majority of agents played $+1$). A strategy 
is a mapping from the $2^m$ possible price histories onto the two possible
decisions. 
An example of a strategy with $m=3$ is
\ba
&& \{000 \to 0; 001 \to 0; 010 \to 1; 011 \to 0;  \nonumber \\
&& 100 \to 1;  101 \to 0; 110 \to 1; 111 \to 0\}~.   
\label{stramfm}
\ea
Different payoffs of strategies define different mechanisms and 
apply to distinct situations. The standard MG corresponds to 
the payoff $g_j(t) = - a_j(t) A(t)$:  if a strategy $j$ is in the minority
($a_j(t)A(t) < 0$), it is rewarded. In other words, agents in the MG try
to be anti-imitative. Another payoff, for instance motivated
by real financial markets, is \cite{$-game,Giardina} $g_j(t) = a_j(t-1) A(t)$. In
this case, the price is an increasing function of the excess demand
$A(t)$ and the payoff reflects the tendency for a strategy to win if it
has anticipated correctly the next market move. Our results apply
equally well to these two as well as other payoffs.   

Given the common information $\mu_m(t)$ at time $t$, we
define the process of decoupling of a strategy as follows: \\
\noindent $\bullet$ A strategy $s_j$
is called $n-$time steps decoupled conditioned on $\mu_m(t)$ if
the action $s_j(\mu_m (t+n+1))$ does not
depend on $\mu_m(t+1), ..., \mu_m(t+n)$. \\
\noindent $\bullet$ A strategy $s_i(\mu_m (t+n+1))$, whose
action at time $t+n+1$ 
depends on at least one of the outcomes of $\mu_m(t+1), ..., \mu_m(t+n)$, is
coupled to the information flow.\\
\noindent As an example, consider the strategy
(\ref{stramfm}). This strategy is one-time step decoupled
conditioned on $\mu_3(t) =000$ or $\mu_3(t) =100$ since, in both 
realisations 
$\mu_3(t+1) =000$ or $\mu_3(t+1) =001$, the strategy's action is $0$ 
at time $t+2$.
More generally, a strategy $s$ with $m=3$ is one-step decoupled
conditioned on the common information $\mu_3(t)=abc$ if
$s(bc0)=s(bc1)$. It is then automatically decoupled conditioned
on $\mu_3(t)={\bar a}bc$ where ${\bar a} = 1$ if $a=0$ and ${\bar a} = 0$ if $a=1$.
The fraction of one-step decoupled strategies conditioned on 
having only one pair $(abc, {\bar a}bc)$ decoupled 
is ${4 \choose 1} 2^{-4}=1/4$, since there are four possible
pairs $(bc)$ each having a probability $1/2$ that $s(bc0)=s(bc1)$.
The strategy $s$ is two-step decoupled conditioned on $\mu_3(t)=abc$ if
$s(c00)=s(c01)=s(c10)=s(c11)$. 
For the general $m$ memory case,
a strategy is one-step decoupled conditioned on $\mu_m(t)=c_1 c_2 ... c_m$
if $s(c_2 ... c_m 0) = s(c_2 ... c_m 1)$.  The fraction of one-step
decoupled strategies on at least one pair of histories is $1-2^{-2^{m-1}}$,
which is one minus the probability that none among the number $2^{m-1}$ of $(m-1)$-plets obey
$s(c_2 ... c_m 0) = s(c_2 ... c_m 1)$.
A strategy is 
$n$-step decoupled (with $n \leq m$) if 
$s(c_{n+1} ... c_m \{\mu_n\})$ is independent of all possible $2^n$ histories
$\{\mu_n\}$ of string length $n$. The fraction of $n$-step 
decoupled strategies conditioned on any 
$(m-n)$-plets only is $1-(1-2^{-(2^n -1)})^{2^{m-n}}$, because
$2^{-(2^n -1)}$ is the probability for a given history to be
$n$-step decoupled, thus $1-2^{-(2^n -1)})$ is the probability 
to be $n$-step coupled and there are $2^{m-n}$ $(m-n)$-plets.
Thus, as soon as $n$ becomes larger than $ \approx \ln_2(m)$, the 
$n$-step decoupled strategies become extremely rare.

What is interesting about decoupled strategies is that they
open the possibility that one can predict with {\em certainty}\cite{note1}
the future global action in two (or more) time steps ahead without
having to know it at the next time step. This
occurs when a majority of agents use decoupled strategies which combine
to a majority action. Indeed, the common action of
all the agents at time $t+n+1$ can be written as the sum of two contributions, 
one stemming from coupled and the other from decoupled strategies
(conditioned on the history ${\mu_m}(t)$):
\be
A^{{\mu_m} (t)} \equiv A_{\rm coupled} ^{{\mu_m} (t)} + A_{\rm
decoupled}^{{\mu_m} (t)} ~.
\label{mngmekl}
\ee
The condition for certain predictability $n$ time steps ahead is therefore
\be
|A_{\rm decoupled }^{{\mu_m} (t)} (t+n+1)| > N/2~.
\label{cojmgfwl}
\ee
We call these times when condition (\ref{cojmgfwl}) is met ``prediction days.''

It is interesting to estimate the 
frequency of such prediction days that would be obtained if strategies
and histories were randomly chosen. Let us first estimate
the probability ${\rm Pr}_1$ 
that a given agent is $n$-step decoupled. This is the probability
that he is active and that his best strategy is $n$-step decoupled. The former
is a fixed finite fraction $0<\eta<1$ of time. The probability that his best
strategy is $n$-step decoupled is the product of the probability that his
best strategy belongs to the set of possibly $n$-step decoupled strategies
times the probability that the present history is decoupling for that strategy.
The former is $1-(1-2^{-(2^n -1)})^{2^{m-n}}$ as shown above.
The later is $1/2^{2^n-1}$, which is the probability for $2^n$
values $s(c_{n+1} ... c_m \{\mu_n\})$ to be equal for all
$2^n$ histories $\{\mu_n\}$. This gives
$$
{\rm Pr}_1(m,n) = \eta \left[ 1-\left(1-{1 \over 2^{2^n-1}}\right)^{2^{m-n}} 
\right] {1 \over 2^{2^n-1}}.  
$$
We have for instance ${\rm Pr}_1(m=3,n=1)=3\eta/8$; ${\rm Pr}_1(m=3,n=2)=11\eta/216$,
${\rm Pr}_1(m\to +\infty,n=1) \to \eta/2$ from below, and
${\rm Pr}_1(m,n) < \eta/2$ for all $n \leq m$.

\begin{figure}
\includegraphics[width=8cm]{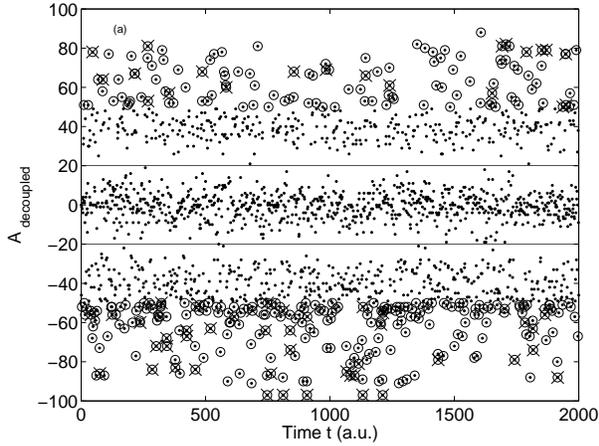}
\caption{\protect\label{Fig1} 
$A_{\rm decoupled}$ defined in (\ref{mngmekl}) 
as a function of time for the MG with $N=101,
s=12, m=3$. Circles indicate one-step prediction days, crosses are
the subset of days starting a run of 
two or more consecutive one-step prediction days.
}
\end{figure}

Assuming complete
incoherence between the strategies and histories, the probability 
${\rm Pr}_{\rm pred}$ that 
condition (\ref{cojmgfwl}) is met is
\be
{\rm Pr}_{\rm pred} = 2 \sum_{D={N \over 2}}^N [{\rm Pr}_1]^D
\sum_{D_+ > {D \over 2}+{N \over 4}}^D {D \choose D_+} {1 \over 2^D}
\label{mgnkmgks}
\ee
where $D$ is the number of active decoupled agents and $D_+$ is
the number among them who take a positive step. The factor $2$ comes 
from the two possible signs of $A_{\rm decoupled }^{{\mu_m} (t)}$.
The lower bound ${D \over 2}+{N \over 4} < D_+$ in the second sum ensures that condition
(\ref{cojmgfwl}) is obeyed. The lower bound $D=N/2$ in the first sum 
expresses the obvious fact that condition (\ref{cojmgfwl})  is met
when the number of decoupled agents is at least $N/2$ (this lower bound
is reached when they all agree). Now, the second sum is obviously bounded from 
above by $1$. This yields 
\be
{\rm Pr}_{\rm pred} < {2 {\rm Pr}_1^{N/2} \over 1- {\rm Pr}_1}
< {4 \over 2-\eta} \left({\eta \over 2}\right)^{N/2} < {4  \over 2^{N/2}}~,
\label{mgnksd}
\ee
where the two last bounds use the fact that ${\rm Pr}_1(m,n) < \eta/2 < 1/2$ for all $n, m$.
This result (\ref{mgnksd}) shows that in this scenario, as soon as one considers populations
of agents of the order of a few tens or more, the probability to
find a prediction day is exceedingly small. In our simulations below, we have
used $N=25$ and $N=101$, which gives respectively ${\rm Pr}_{\rm pred} 
< 7 \cdot 10^{-4}$ and  ${\rm Pr}_{\rm pred} < 2.5 \cdot 10^{-15}$.

Figure \ref{Fig1} shows a typical time series of $A_{\rm decoupled}$ 
as a function of time for the standard MG with $N=101, s=12, m=3$. 
When strategies are randomized at each time step, there are no prediction day
as espected from our above estimated probability (\ref{mgnksd})
which applies to this case and $A_{\rm decoupled}$ remains
confined around zero within the strip delineated by the two horizontal lines
at $\pm 20$. In constrast, Fig.~\ref{Fig1} shows a surprisingly large fraction 
$\rho_{\rm pred}=17.5\%$ for one-step prediction days
and $3.5\%$ for two or more consecutive days which are predictive.
This implies that the prediction days result from a collective
organization of agents and of their strategies which condense into a herding regime
characterized by a rather strong synchronization of both their decoupling and of
their action conditioned on decoupling.

\begin{figure}
\includegraphics[width=8cm]{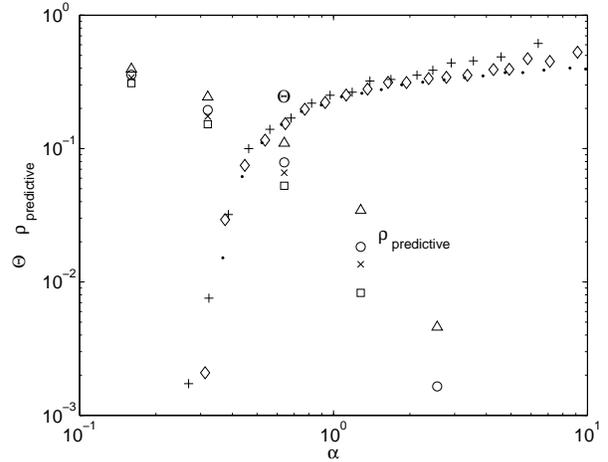}
\caption{\protect\label{Fig2}
$\theta \equiv {\overline{\langle {\rm sign}[A] \rangle^2}}$ and
the frequency of predictive days, $\rho_{\rm pred}$ versus
$\alpha \equiv 2^m/N$. Simulations for $\theta$ were done with
$s=2$, $m=5$ ($+$), $m=6$ (diamonds), $m=7$ (black dots). Simulations for
$\rho_{\rm predictive}$ were done with $N=25$, $s=7$ (squares),
$s=8$ (crosses), $s=9$ (circles), $s=12$ (triangles).
}
\end{figure}

By scanning different values
for the number $s$ of strategies per agent and for the memory $m$,
figure \ref{Fig2} shows that the new phenomenon discovered here is
a robust property of MG.
We also compare the occurrence of our prediction days with a 
measure of predictability $\theta \equiv {\overline{\langle {\rm sign}[A^{\mu}]\rangle^2}}$
introduced by Savit et al. \cite{CM},
where $\langle {\rm sign}[A^{\mu}]\rangle$ is the time-average of the
collective action conditioned upon the occurrence of a given history $\mu$, while
the upper bar denotes the average over all possible histories. Previous 
works (see also figure \ref{Fig2}) have shown that there
is a critical value $\alpha_c \approx 0.35$ (where $\alpha \equiv 2^m/N$) such that 
$\theta$ goes from non-zero (statistically predictable regime) for $\alpha > \alpha_c$
to zero (statistically unpredictable regime) for $\alpha < \alpha_c$. Our
simulations illustrate a case where our prediction days are in fact more frequent
in the previously classified ``statistically unpredictable'' regime, showing
that we are dealing with a fundamentally different property.
Note that $\rho_{\rm pred}$ grows as $\alpha$ decreases, i.e., when $m$ decreases
and $N$ increases, in blatant contradiction with the expectations
(\ref{mgnkmgks}) assuming incoherent and random choices.
Again, this reinforces the evidence that the prediction days results
from a special herding organization of the strategies in conjunction with the 
realized histories. 

\begin{figure}
\includegraphics[width=8cm,width=8cm]{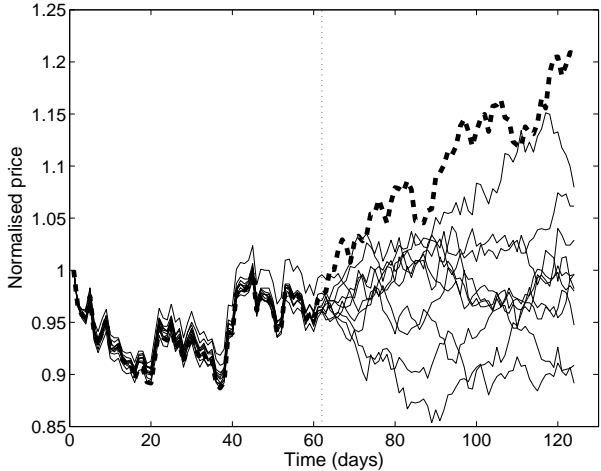}
\caption{\protect\label{Fig3} Fat dashed line:
Nasdaq Composite price history (black-box game) as 
a function of time (days); thin solid lines:
ten predicted price trajectories obtained from the third-party games.
The first in-sample 61 days are used to calibrate ten third-party games. The days
62-123 are out-of-sample. The third-party games make a poor job at predicting the 
out-of-sample prices of the Nasdaq Composite index, while table 1 shows that 
they predict specific pockets of predictability
associated with forecasted ``prediction days'' (see text).
}
\end{figure}

A major objection could be raised that 
the information required for identifying a prediction day 
includes, not only the common knowledge of past history but, the
active strategies used by the agents, which are in general unobservable.
Actually, it is possible to retrieve
sufficient information on the strategies by
using a methodology derived from Lamper et al.
\cite{Lamper}, which allows one to invert a time series of 
observable collective action $A(t)$
generated by a black-box game to obtain
an ensemble of so-called third-party games by
optimizing a measure of the matching between the observed time series and synthetic
time series generated with the third-party games \cite{fut}.
It is then possible to identify
the prediction days in the third-party games. To quantify 
how these prediction days by third-party games provide a forecast for the black-box
game, we use as a metric the success rate to forecast the correct sign of the
global action $A$ on a prediction day identified by the third-party games
compared with the success rate on other days. In particular, we monitor
how these success rates vary with the predicted amplitude of 
$A_{\rm decoupled}$, since expression (\ref{mngmekl}) shows that
the larger it is, the more predictable is the global action.
We find that, for both the MG
as well as for the majority \$-game \cite{$-game}, the former success rate shows
the predicted behavior of increasing monotonically from 50\% for small
$A_{\rm decoupled}$ to 100\% for large $A_{\rm decoupled}$. In contrast, 
we do not observe any significant prediction skill above 50\% on non-prediction days.
Rather than reporting these results in details on synthetic MG and \$ games, we now
show that our method works for real complex system of competing agents, and we
take the stock market as a significant application. We use the return
time series of the Nasdaq Composite index as the proxy for the global
action $A(t)$, whose price trajectory is shown in figure \ref{Fig3}. We
construct ten third-party \$-games (with the same parameters
but different realizations of strategies endowed to the agents) 
calibrated on the first $61$ daily returns
to the left of the vertical dashed line in figure \ref{Fig3}. We then feed 
to these third-party games the realized returns of the Nasdaq Composite and 
compare their predictions with the realized price. The predictions are issued
at each close of the day for the next close, all third-party games being unchanged
for the test over the second part of the time period from $t=62$ to $123$.
Among the $62$ out-of-sample days, we monitor our ten third-party games
to detect a prediction day (we use a voting process among the ten third-party
games to obtain better robustness), according to the active strategies
and the predicted one-step ahead history for the next close. Conditioned
on the detection of a prediction day, we issue a prediction of the sign of the
next day return and compare it with the realized market return. 
The performance of this prediction scheme is reported in table 1 which is 
typical of our results. The important point is that the success rate ($\%$
in the table) increases
with the predicted amplitude of $A_{\rm decoupled}$, as for the 
synthetic MG and \$-game mentioned above.

\begin{table}
\begin{center}
\begin{tabular} {|c|c|c|c|c|c|c|c|c|c|c|} \hline \hline
$|A|$ & 0 & 0.5 & 1 & 1.5 & 2 & 2.5 & 3 & 3.5 & 4 & 4.5  \\
\% & 53 & 61 & 67 & 65 & 82 & 70 & 67 & 67 & 100 & 100   \\
{\rm Nb} & 62 & 49 & 39 & 23 & 17 & 10 & 6 & 3 & 2 & 1  \\ \hline \hline
\end{tabular}
\end{center}
\caption{Out-of-sample success rate \% (second row) using different thresholds
for the predicted global decoupled action (first row) of the 
third-party \$-games calibrated to the Nasdaq Composite index. ${\rm Nb}$ (third row)
is the number of days from $t=62$ to $123$ which have their predicted global
decoupled action $|A_{\rm decoupled}|$ larger than the value indicated in the first row.}
\end{table}

In contrast, the use of the third-party games for predicting each day the sign
of the next return fails, as can be seen from the ten trajectories in the out-of-sample
time interval. Our method has thus identified pockets of predictability 
in the Nasdaq index associated with
the prediction of ensembles of decoupled strategies predicted by the third-party games.

{}


\begin{thebibliography}{}
\bibitem{note1}
Ignoring the indeterminacy that comes from strategies that 
are degenerate in score as well ignoring any change in score 
between time $t$ and $t+n$.
\bibitem{fut} The details of the method will be reported elsewhere.
\end{thebibliography}
\end{document}